\documentclass{statsoc}

\usepackage[a4paper]{geometry}
\usepackage{graphicx}
\usepackage{amsmath, amssymb}
\usepackage{natbib}
\usepackage{color,bm,xcolor}
\usepackage{enumitem}
\usepackage{multirow}
\usepackage{etoolbox}

\makeatletter
\patchcmd{\@makecaption}
  {\parbox}
  {\advance\@tempdima-\fontdimen2} 
  {}{}
\makeatother  

\def\bSig\mathbf{\Sigma}

\newcommand{\Bin}{\text{Bin}}
\newcommand{\E}{\text{E}}
\newcommand{\N}{\text{N}}
\newcommand{\logit}{\text{logit}}

\newcommand{\data}{\text{Data}}
\newcommand{\Corr}{\text{Corr}}

\providecommand{\keywords}[1]{\textbf{\textit{Keywords:}} #1}
\newcommand{\iid}{\stackrel{iid}{\sim}}

\makeatletter
\newcommand{\pushright}[1]{\ifmeasuring@#1\else\omit\hfill$\displaystyle#1$\fi\ignorespaces}
\newcommand{\pushleft}[1]{\ifmeasuring@#1\else\omit$\displaystyle#1$\hfill\fi\ignorespaces}
\makeatother

\title[MUCE: A Bayesian Design for Phase 1b Multiple Expansion Cohort Trials]{MUCE: Bayesian Hierarchical Modeling for the Design and Analysis of Phase 1b Multiple Expansion Cohort Trials}
\author{Jiaying Lyu\thanks{These authors contributed equally}}
\address{Laiya Consulting, Inc., Chicago, USA}
\author{Tianjian Zhou\footnotemark[1]}
\address{University of Chicago, USA}
\author{Shijie Yuan, Wentian Guo}
\address{Laiya Consulting, Inc., Chicago, USA}
\author[J. Lyu, T. Zhou, S. Yuan, W. Guo and Y. Ji]{and Yuan Ji}
\address{University of Chicago, USA}
\email{yji@health.bsd.uchicago.edu}

\begin{document}
\begin{abstract}
We propose a \underline{mu}ltiple \underline{c}ohort \underline{e}xpansion (MUCE) approach as a design or analysis method 
for phase 1b multiple expansion cohort trials, which are novel first-in-human studies conducted following phase 1a dose escalation.
In a phase 1b expansion cohort trial, one or more doses of a new investigational drug identified from phase 1a are tested for initial anti-tumor activities in patients with different indications (cancer types and/or biomarker status). Each dose-indication combination defines an arm, and patients are enrolled in parallel cohorts to all the arms. 
The MUCE design is based on a class of Bayesian hierarchical models that adaptively borrow information across arms. Specifically, we employ a latent probit model that allows for different degrees of borrowing across doses and indications.
Statistical inference is directly based on the posterior probability of each arm being efficacious, facilitating the decision making that decides which arm to select for further testing. 
The MUCE design also incorporates interim looks, based on which the non-promising arms will be stopped early due to futility.
Through simulation studies, we show that MUCE exhibits superior operating characteristics. We also compare the performance of MUCE with that of Simon's two-stage design and existing Bayesian designs for multi-arm trials.
To our knowledge,  MUCE is the first Bayesian method
for phase 1b expansion cohort trials with multiple doses and indications. 
\end{abstract}

\section{Introduction}
\label{sec:intro}

\begin{figure}[h!]
\centering \includegraphics[width=0.95\textwidth]{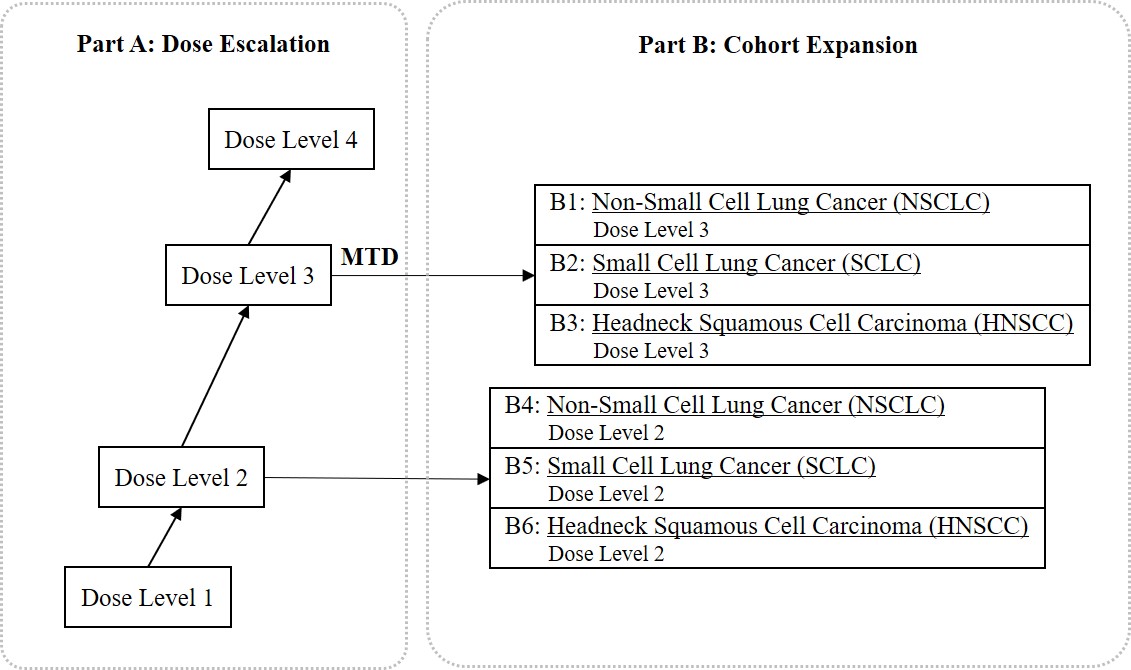}
\caption{A stylized depiction of a two-part phase 1 study. Part A is the phase 1a dose escalation and Part B is the phase 1b cohort expansion. In the dose-escalation stage, four candidate dose levels are investigated for safety, and dose level 3 is identified as the maximum tolerated dose (MTD), i.e., the highest dose with tolerable side effects.
In the cohort-expansion stage, the MTD and the dose  below (dose level 2) are considered for further investigation. Patients with three different indications are enrolled in parallel for both doses.
This leads to a total of six cohorts, cohorts B1--B3 for dose level 3 and cohorts B4--B6 for dose level 2. } \label{fig:phase-1}
\end{figure}

Phase 1b expansion cohort trials are a relative new type of studies to investigate the anti-tumor effects of multiple doses of a new treatment in multiple indications. Here, indications can be different cancer types according to histology, biomarker status, or both. Figure \ref{fig:phase-1} is a stylized depiction of a phase 1a/1b trial for a new drug: Part A refers to the phase 1a dose escalation in which different dose levels of the drug are investigated for safety; Part B is the phase 1b cohort expansion stage, in which one or more candidate dose levels with reasonable safety profiles are selected for further evaluation of efficacy. Both parts can be incorporated seamlessly in a single design or separated as two different trials, depending on the practical situation for each drug development. In phase 1b, patients with different indications are enrolled in parallel to these candidate doses,  and their efficacy outcomes are recorded. Since multiple doses and multiple indications may be tested, we refer to a dose-indication combination as an ``arm'', 
e.g., arms B1--B6 in Figure \ref{fig:phase-1}. 
At the end, the response rate of an arm is estimated, based on which a go/no-go decision about further development of the dose and indication is made.

In 2018, the U.S. Food and Drug Administration released a draft guidance \citep{us2018expansion} that recommends the use of  multiple expansion cohort  trials to expedite oncology drug development. A statistical design  mentioned in this draft guidance is the Simon's two-stage design \citep{simon1989}. The Simon's two-stage design provides trial sample size calculation and trial conduct for a binary endpoint (efficacy response/no response) under the hypothesis test of  $H_{0k}: p_{k} \leq \pi_{k0}$ versus $H_{1k}: p_{k} \geq \pi_{k1}$. Here, $k$ is a specific arm in the phase 1b trial, $p_{k}$ is the objective response rate (ORR) in arm $k$, $\pi_{k0}$ is the reference response rate, such as the rate under the standard of care (SOC), and $\pi_{k1}$ is the target response rate, with which the drug is regarded superior. An objective response refers to a partial or complete response according to the RECIST guideline \citep{eisenhauer2009}. 
The Simon's two-stage design proceeds as follows: in the first stage, $n_1$ patients are enrolled, and the trial is stopped if $r_1$ or fewer patients respond. Otherwise, additional $(N-n_1)$ patients are enrolled in the second stage. The drug is considered promising and $H_0$ is rejected if more than $r$ patients (including those from the first stage) respond. The tuple $(r_1, n_1, r, N)$ is determined based on the desired control of frequentist type I and II error rates and certain optimality conditions, such as minimizing the expected sample size.  The Simon's two-stage design is appealing for single-arm studies, since the design can limit the number of patients exposed to an inefficacious drug.
To apply the Simon's two-stage design to multi-arm trials, one could treat each arm as a separate study, and the tuple $(r_1, n_1, r, N)$ is determined for each arm under arm-specific type I and type II error rates.
The BRAF-V600 study in \cite{hyman2015} is an example of using the Simon's two-stage design in a multi-arm trial.
However, the Simon's two-stage design was developed for single-arm trials and may not be the most efficient design for multi-arm trials (including multiple expansion cohort trials) for at least two reasons.
First, an important rationale to include multiple arms (e.g., multiple doses and indications) into a single study is
that the treatment effects in some arms may provide information about the treatment effects in other arms.
Therefore, it is desirable to borrow information across arms when we design the trial and perform data analysis. Second, in a multi-arm trial, applying the Simon's two-stage design independently to each arm does not take into account the family-wise type I error rate (FWER). For example, consider a trial with 4 arms. Suppose each arm is designed with a type I error rate of 0.1, then the FWER can be as high as $1-(1-0.1)^4 = 0.35$, which means that with a probability of $0.35$ an inefficacious arm may be recommended for further development. 
Of course, to guarantee that the FWER is no higher than $\alpha$, one could apply the Bonferroni correction and require the type I error rate for each arm to be no higher than $\alpha / K$, where $K$ is the number of arms. However, that may result in  a large sample size for early-phase trials.

Several Bayesian designs have been proposed for multi-arm clinical trials, such as \cite{thall2003hierarchical}, \cite{berry2013bayesian}, \cite{neuenschwander2016robust}, \cite{simon2016bayesian}, \cite{liu2017increasing}, \cite{cunanan2017efficient}, \cite{chu2018bayesian}, \cite{chu2018blast}, \cite{hobbs2018bayesian} and \cite{psioda2019bayesian}, among others.
A majority of these designs make use of Bayesian hierarchical models to borrow information across arms and increase statistical efficiency.
Most of these designs are developed for basket trials \citep{heinrich2008,menis2014,hyman2015}, which evaluate a new treatment in multiple indications (without the notion of multiple doses).
In this paper, we extend the idea of existing Bayesian designs for multi-arm trials and develop a design specifically for multiple expansion cohort trials.
The proposed design is called MUCE, which stands for \underline{mu}ltiple \underline{c}ohort \underline{e}xpansion.
A unique feature of multiple expansion cohort trials is that they could have a two-dimensional dependency structure across doses and indications.
For example in Figure \ref{fig:phase-1}, when two  doses are expanded in three indications, dependence in both doses and indications may affect model performance.

\paragraph{A motivating example.}
We introduce a case study that motivates the MUCE design.
Consider a seamless phase 1a/1b trial that evaluates the safety and efficacy of a bispecific monoclonal cancer drug.
In phase 1a, five doses are tested for safety. The endpoint is dose-limiting toxicity (DLT). Phase 1a is guided by the i3+3 design \citep{liu2019design}, which employs a set of rules to make dose-escalation decisions. The maximum tolerated dose (MTD) will be identified from phase 1a dose finding. Up to three doses, none higher than the MTD, will be considered for expansion in the phase 1b study, and four different indications based on histology will be considered. This leads to a maximum of 12 arms, each with a unique dose-indication combination. The phase 1b endpoint is objective response. The ORR of each arm is compared to a historical rate. Specifically, for arm $k$, we intend to test the null hypothesis $H_{0k}: p_{k} \leq \pi_{k0}$ versus the alternative hypothesis $H_{1k}: p_{k} > \pi_{k0}$, with $\pi_{k0} = 0.2$ being the historical response rate for all arms.
The Simon's two-stage design may not be the best choice for this trial. To see this, consider applying the Simon's two-stage design independently to each arm.
In the extreme case, 12 arms will be expanded in phase 1b. Then, the Simon's two-stage design with an arm-specific type I error rate of $\alpha = 0.1$ would result in a FWER of  $1-(1 - 0.1)^{12} = 0.72$. Apparently, this is not acceptable since such a FWER would render a great risk for downstream clinical development. 
In addition, the trial budget only allows about 10 patients per arm, making it difficult to use the Simon's two-stage design with decent power. It is important to borrow information to allow reasonable power. We will present numerical results for this trial based on the MUCE design later.

Motivated by the case study, we develop the MUCE design to power multi-dose-indication expansion cohort trials. 
The MUCE design is based on a class of Bayesian hierarchical models that allows different degrees of borrowing across the two dimensions -- doses and indications.
For example, the drug may perform more similarly across different doses than across different indications. 
This is different from existing Bayesian designs for multi-arm trials, which were developed aiming for one-dimensional borrowing.
In addition, the MUCE design makes inference directly based on the posterior probability of the alternative hypothesis $\Pr(H_{1k} \mid \data)$.
Through a Bayesian hierarchical model including prior probabilities of the hypotheses, we follow the argument in \cite{scott2010bayes} to realize Bayesian multiplicity control.
We will demonstrate through simulation studies that the MUCE design has desirable operating characteristics.

The remainder of the paper is organized as follows. In Section \ref{sec:rev}, we provide a brief review of existing Bayesian designs for multi-arm trials. In Section \ref{sec:method}, we propose the probability model and the decision rules for the MUCE design. 
In Section \ref{sec:result}, we evaluate the operating characteristics of the proposed MUCE design and present simulation results. The paper concludes
with a discussion in Section \ref{sec:discussion}.

\section{Review of Bayesian Designs for Multi-arm Trials}
\label{sec:rev}

In this section, we provide an overview of existing Bayesian designs for multi-arm trials.
Let $K$ denote the number of arms in the trial.
For example, in a multiple expansion cohort trial, $K$ is the number of arms, i.e., dose-indication combinations; in a basket trial, $K$ is the number of indications.
Let $n_k$ and $y_k$ denote the number of patients and responders in arm $k$, respectively.
Here, a responder refers to a cancer patient who has a beneficial outcome after the treatment, where the beneficial outcome is usually defined based on tumor shrinkage or some meaningful anti-tumor activities (e.g., those in \citealp{eisenhauer2009}).
Denote by $p_k$ the true and unknown response rate for arm $k$. 
A natural sampling model for $y_k$ is the binomial model, $y_k \mid p_k \sim \Bin(n_k, p_k)$.

\cite{berry2013bayesian} propose a Bayesian hierarchical model (BBHM) which  borrows strength across different arms. 
For each arm $k$, BBHM considers a hypothesis test:
\begin{align*}
H_{0k}: p_k \leq \pi_{k0} \quad \text{versus} \quad H_{1k}: p_k \geq \pi_{k1},
\end{align*}
where $\pi_{k0}$ and $\pi_{k1}$ are the reference and target response rates for arm $k$, respectively.
Let $\theta_k = \logit(p_k) - \logit(\pi_{k1})$ denote the log-odds of the response rate including an adjustment for the targeted rate $\pi_{k1}$, where $\logit(x) = \log[x / (1-x)]$.
BBHM models the $\theta_k$'s via a shrinkage prior given by
\begin{align}
\theta_k \mid \theta, \sigma^2 \iid \N(\theta, \sigma^2), \quad k = 1, \ldots, K.
\label{eq:theta_k}
\end{align}
The hyperparameters $\theta$ and $\sigma^2$ are given conjugate hyperpriors,
\begin{align*}
\theta \sim  \N(\theta_0,\sigma^2_0), \quad \text{and} \quad
\sigma^2 \sim  \text{Inv-Gamma}(\alpha_0,\beta_0).
\end{align*}
This prior construction assumes that the $\theta_k$'s
across different arms are exchangeable and are shrunk toward a shared mean $\theta$, which enables borrowing information across the $K$ arms. The degree of borrowing is determined by the value of $\sigma^2$. The smaller the $\sigma^2$, the stronger the borrowing. 
On one extreme, when $\sigma^2=0$, all the $\theta_k$'s are equal, resulting in full shrinkage. 
On the other extreme, when $\sigma^2$ goes to infinity, the degree of borrowing goes to zero. 

The BBHM design incorporates interim analyses for futility stopping. Specifically, each interim analysis occurs after a pre-specified number of patients are enrolled.
If 
\begin{align}
\Pr \left( p_k > \frac{\pi_{k0}+\pi_{k1}}{2} \mid \text{Interim data} \right) < {\phi_1},
\label{eq:interval-futility}
\end{align}
enrollment to arm $k$ is stopped for futility; otherwise, enrollment to arm $k$ continues until the next interim analysis or the end of the trial.
At the end of the trial, a final analysis is conducted, and arm $k$ is declared efficacious and promising for further study if 
\begin{align}
\Pr(p_k > \pi_{k0} \mid \text{Final data})> {\phi_2}.
\label{eq:interval}
\end{align}
Here, $\phi_1$ and $\phi_2$ are tuning parameters, which may be determined through simulation studies to generate desirable frequentist operating characteristics.

BBHM design shows superior power when most arms are truly efficacious. The cost is the inflated frequentist type I error rates for non-promising arms, if the degree of borrowing is overestimated.
See, e.g., \cite{neuenschwander2016robust},  \cite{chu2018bayesian} and \cite{chu2018blast} for discussions.
Several alternative methods have been proposed, attempting to mitigate the issue.
\cite{neuenschwander2016robust} propose the exchangeability-nonexchangeability (EXNEX) design, which models the $\theta_k$'s in Equation \eqref{eq:theta_k} with a mixture distribution,
\begin{align*}
\theta_k  \sim \sum_{c = 1}^C w_{kc} \N(\theta_{\text{EX}, c}, \sigma_{\text{EX}, c}^2) + w_{k0} \N(\theta_{\text{NEX}, k}, \sigma_{\text{NEX}, k}^2).
\end{align*}
In other words, with probability $w_{kc}$, $\theta_k$ belongs to an exchangeability (EX) component $c$, and with probability $w_{k0}$, $\theta_k$ belongs to a nonexchangeability (NEX) component;  $\sum_{c = 0}^C w_{kc} = 1$.
The parameters of the EX components, $\theta_{\text{EX}, c}$ and $\sigma_{\text{EX}, c}$, are shared across arms within component $c$. In contrast, the parameters of the NEX components, $\theta_{\text{NEX}, k}$ and $\sigma_{\text{NEX}, k}$, are arm-specific.
The number of EX components $C$ and the weights of the components $\bm w_k = (w_{k1}, \ldots, w_{kC}, w_{k0})$ are prespecified. 
The authors recommend as a default setting that the same NEX components and mixture weights are used for all arms, i.e., $\theta_{\text{NEX}, 1} = \ldots = \theta_{\text{NEX}, K} = \theta_{\text{NEX}}$, $\sigma^2_{\text{NEX}, 1} = \ldots = \sigma^2_{\text{NEX}, K} = \sigma^2_{\text{NEX}}$, and $\bm w_1 = \ldots = \bm w_K = \bm w$. The NEX variance $\sigma_{\text{NEX}}^2$ should be chosen large to ensure a good performance. 
Interestingly, this default setting collapses all the nonexchangeable components into a single component, effectively rendering the model ``exchangeable''. 
However, the use of the mixture model in EXNEX reduces the extent of borrowing across arms thus leads to less type I error inflation compared to BBHM. 
The original EXNEX design does not have a futility stopping rule, but the same rule as in Equation \eqref{eq:interval-futility} may be included.

\cite{chu2018bayesian} propose a calibrated Bayesian hierarchical model (CBHM), which uses an empirical Bayes 
estimate of $\sigma^2$ in Equation \eqref{eq:theta_k} rather than placing a prior on it. 
This calibration process results in more conservative estimation of $\sigma^2$ compared to BBHM when the treatment effects in different arms are less homogeneous, leading to less borrowing and type I error inflation.
The CBHM design has the same decision rules for futility stopping and declaring efficacy as the BBHM design.

\section{The MUCE Design}
\label{sec:method}

The MUCE design takes a slightly different angle.
Instead of using the posterior credible interval of the estimated response rate (Equations \ref{eq:interval-futility} and \ref{eq:interval}) for decision and inference, in MUCE we propose a hierarchical model incorporating the hypotheses as a parameter, i.e., Bayesian hypothesis testing.
Also, to exploit the data structure in multiple expansion cohort trials, we construct a latent probit model that allows different degrees of borrowing across doses and indications.
This will be more clear in the upcoming discussion.

\subsection{Probability Model}
\label{sec:prob_model}

Consider a phase 1b trial that evaluates $J$ different dose levels of a new drug in $I$ different indications.
Let $(i, j)$ denote the arm for indication $i$ and dose level $j$, $i = 1, \ldots, I$, $j = 1, \ldots, J$.
The total number of arms is $K = I \times J$.
Suppose $n_{ij}$ patients have been treated in arm $(i, j)$, and $y_{ij}$ of them are responders.  Let $p_{ij}$ denote the true and unknown response rate for the arm $(i, j)$.  We assume $y_{ij}$ follows a binomial distribution, 
$y_{ij} \mid p_{ij} \sim \Bin(n_{ij},p_{ij})$.
Whether dose level $j$ is efficacious for indication $i$ can be examined by the following hypothesis test:
\begin{align}
H_{0,ij}: p_{ij}\leq \pi_{i0} \quad \text{versus} \quad
H_{1,ij}: p_{ij} > \pi_{i0},
\label{eq:test}
\end{align}
where $\pi_{i0}$ is the reference response rate for indication $i$.
For simplicity, we do not separately consider a target response rate $\pi_{i1}$ as in the Simon's two-stage and BBHM designs. This is because only the reference response rate is used for declaring treatment efficacy in the final analysis for all the existing Bayesian designs (Equation \ref{eq:interval}).

Under a formal Bayesian testing framework for \eqref{eq:test}, let $\lambda_{ij}$ be a binary and random indicator of the hypothesis, such that $\lambda_{ij} = 0$ (or 1) represents that hypothesis $H_{0,ij}$ (or $H_{1,ij}$) is true. 
We formally construct a hierarchical model treating $\lambda_{ij}$ as a model parameter and perform inference on $\lambda_{ij}$ directly.
In the first step, we build a prior model for $p_{ij}$ under each hypothesis. Similar to BBHM, we consider the logit transformation of $p_{ij}$, $\theta_{ij}=\logit(p_{ij})$. The null hypothesis $p_{ij} \leq \pi_{i0}$ is equivalent to $\theta_{ij} \leq \theta_{i0}$, and the alternative hypothesis is equivalent to $\theta_{ij} > \theta_{i0}$, where $\theta_{i0}=\logit(\pi_{i0})$. Conditional on $\lambda_{ij}$, we assume
\begin{align*}
\theta_{ij} \mid \lambda_{ij} = 0 &\sim \text{Trunc-Cauchy}(\theta_{i0}, \gamma; (-\infty, \theta_{i0}]), \\
\theta_{ij} \mid \lambda_{ij} = 1 &\sim \text{Trunc-Cauchy}(\theta_{i0}, \gamma; (\theta_{i0}, \infty)),
\end{align*}
where $\text{Trunc-Cauchy}(\theta, \gamma; A)$ denotes a Cauchy distribution with location $\theta$ and scale $\gamma$ truncated within interval $A$. The use of the Cauchy distribution priors follows \cite{gelman2008weakly} due to its heavy tail, thus inducing large prior variability and less prior influence.

In the second step, we construct prior models for the probabilities of the hypotheses, $\Pr(\lambda_{ij} = 1)$ and $\Pr(\lambda_{ij} = 0)$. To borrow strength across dose levels and indications, we construct a hierarchical prior model for $\lambda_{ij}$.
A natural and conventional Bayesian approach is to impose a common prior for the probability of $\{\lambda_{ij}=1\}$ (e.g., similar to the prior in Equation \ref{eq:theta_k}), which shrinks the probabilities to a common value.
To better exploit the data structure in multiple expansion cohort trials, we propose to differentiate the borrowing strength from two factors: dose and indication. 
For example, two arms with the same indication or dose might exhibit more similar treatment effects than two arms with different indications and doses.
To achieve this, we use a latent probit two-way ANOVA prior. Let $Z_{ij}$ be a latent Gaussian random variable, and $\lambda_{ij} = I(Z_{ij} \geq 0)$, where $I(\cdot)$ is an indicator function.
Hence $\Pr(\lambda_{ij} = 1) = \Pr(Z_{ij} \geq 0)$.
We model
\begin{align*}
Z_{ij} \sim N({\xi_{i}+\eta_{j}},\sigma_0^2).
\end{align*}
Here, $\E(Z_{ij}) = \xi_i + \eta_j$, in which $\xi_i$ characterizes the  effect of indication $i$ and $\eta_j$ of dose $j$. The indication-specific effects and dose-specific effects are then separately modeled by common priors,
\begin{align*}
\xi_i \mid \xi_0, \sigma_\xi \iid N(\xi_0,\sigma_{\xi}^2), 
\quad \mbox{and} \quad 
\eta_j \mid \eta_0, \sigma_\eta \iid N(\eta_0,\sigma_{\eta}^2).
\end{align*}
Lastly, we put hyperpriors on $\xi_0$ and $\eta_0$, $\xi_0 \sim N(\mu_{\xi_0},\sigma_{\xi_0}^2)$ and $\eta_0 \sim N(\mu_{\eta_0},\sigma_{\eta_0}^2)$.

The entire hierarchical model is summarized in the following display:
\begin{alignat}{2}
&\text{Likelihood:}
&&y_{ij} \mid n_{ij}, p_{ij} \sim  \Bin(n_{ij}, p_{ij}); \nonumber \\
&\text{Transformation:}
&&\theta_{ij}  = \logit(p_{ij}), \theta_{i0} = \logit(\pi_{i0});
\nonumber  \\ 
&\text{Prior for $(\theta_{ij} \mid \lambda_{ij})$:} 
&&\theta_{ij} \mid \lambda_{ij} = 0 \sim \text{Trunc-Cauchy}(\theta_{i0}, \gamma; (-\infty, \theta_{i0}]),
\nonumber \\
&
&&\theta_{ij} \mid \lambda_{ij} = 1 \sim \text{Trunc-Cauchy}(\theta_{i0}, \gamma; (\theta_{i0}, \infty));
\nonumber \\
&\text{Prior for $\lambda_{ij}$:} 
&&\lambda_{ij} = 
\begin{cases}
0,  \quad \text{if $Z_{ij} < 0$}, \\
1,   \quad \text{if $Z_{ij} \geq 0$};
\end{cases}
\label{eq:bhm} \\
&\text{Latent probit regression:} 
&&Z_{ij} \mid \xi_{i}, \eta_{j}, \sigma_{0}^2  \sim  N(\xi_{i}+\eta_{j},\sigma_{0}^2); \nonumber \\
&\text{Indication-specific effects:} \qquad
&&\xi_{i} \mid \xi_0,\sigma_{\xi}^2  \sim  N(\xi_0,\sigma_{\xi}^2); \nonumber \\
&\text{Dose-specific effects:} 
&&\eta_{j} \mid \eta_0,\sigma_{\eta}^2 \sim N(\eta_0,\sigma_{\eta}^2); \nonumber  \\
&\text{Hyperpriors:} 
 &&\xi_{0} \mid \mu_{\xi_0},\sigma_{\xi_0}^2  \sim  N(\mu_{\xi_0},\sigma_{\xi_0}^2), \nonumber \\
& &&\eta_{0} \mid \mu_{\eta_0},\sigma_{\eta_0}^2 \sim N(\mu_{\eta_0},\sigma_{\eta_0}^2). \nonumber 
\end{alignat}
The values of the hyperparameters $\gamma$, $\mu_{\xi_0}$, $\mu_{\eta_0}$, $\sigma_{0}^2$, $\sigma_{\xi}^2$, $\sigma_{\eta}^2$, $\sigma_{\xi_0}^2$ and $\sigma_{\eta_0}^2$ are 
fixed, and the specification of these hyperparameters will be discussed next.

Under the proposed hierarchical model, different $Z_{ij}$'s are \textit{a priori} correlated, thus the model borrows information across arms. To see this, consider the prior correlations of $(Z_{ij}, Z_{i'j'})$ in the following three cases:
\begin{alignat}{2}
&\text{(I) Same indication ($i = i'$):}  &&\Corr(Z_{ij},Z_{ij^{\prime}}) = \frac{\sigma_{\xi}^2+(\sigma_{\xi_0}^2+\sigma_{\eta_0}^2)}{\sigma_0^2+\sigma_{\xi}^2+\sigma_{\xi_0}^2+\sigma_{\eta}^2+\sigma_{\eta_0}^2}, \nonumber\\
&\text{(II) Same dose ($j= j'$):} &&\Corr(Z_{ij},Z_{i^{\prime}j})=\frac{ \sigma_{\eta}^2+(\sigma_{\xi_0}^2 +\sigma_{\eta_0}^2)}{\sigma_0^2+\sigma_{\xi}^2+\sigma_{\xi_0}^2+\sigma_{\eta}^2+\sigma_{\eta_0}^2}, \label{eq:corr}\\
&\text{(III) Different indication \& dose:} \qquad &&\Corr(Z_{ij},Z_{i^{\prime}j^{\prime}}) =\frac{( \sigma_{\xi_0}^2+\sigma_{\eta_0}^2 )}{\sigma_0^2+\sigma_{\xi}^2+\sigma_{\xi_0}^2+\sigma_{\eta}^2+\sigma_{\eta_0}^2}. \nonumber
\end{alignat}
We can see that the degree of borrowing is determined by the relative magnitude of $\sigma_{0}^2$, $\sigma_{\xi}^2$, $\sigma_{\eta}^2$, $\sigma_{\xi_0}^2$ and $\sigma_{\eta_0}^2$, 
with the correlation being the smallest for case (III). For the other two cases, if $\sigma_{\xi}^2>\sigma_{\eta}^2$ (or $\sigma_{\xi}^2<\sigma_{\eta}^2$), the correlation for case (I) is larger (or smaller) than the correlation for case (II), respectively. 
By default, we set $\sigma_{0}^2 = 1$. We will show sensitivity analyses in which desirable degrees of borrowing  could be realized with different choices of variance values.

Lastly, the values of $\mu_{\xi_0}$ and $\mu_{\eta_0}$ affect the prior probability of $\Pr(\lambda_{ij} = 1)$. In particular, more negative values of $\mu_{\xi_0}$ and $\mu_{\eta_0}$ make the prior $\Pr(\lambda_{ij} = 1)$ smaller, and hence the posterior $\Pr(\lambda_{ij} = 1 \mid \text{Data})$ is also smaller given the same likelihood.
We will show that this feature is useful in calibrating MUCE to make it conservative or not in practice, thereby controlling type I error.

\subsection{Posterior inference}
Let $\bm{\theta}$, $\bm{Z}$, $\bm{\xi}$, $\bm{\eta}$ be the set of all $\theta_{ij}$'s, $Z_{ij}$'s, $\xi_{i}$'s and $\eta_j$'s, respectively.
The joint posterior  distribution of the parameters is given by 
\begin{multline*}
p(\bm{\theta},\bm{Z},\bm{\xi},\bm{\eta}, \xi_0, \eta_0 \mid \bm{y}, \bm{n}) 
\propto
\left\{ \prod_{i,j} f(y_{ij}\mid n_{ij}, \theta_{ij}) \cdot \pi(\theta_{ij} \mid Z_{ij})\cdot \pi(Z_{ij}\mid \xi_i,\eta_j) \right\} \cdot \\
\left\{ \prod_{i}\pi(\xi_i \mid \xi_0)\right\} \cdot \left\{ \prod_{j} \pi(\eta_j \mid \eta_0)\right\}  \cdot \pi(\xi_0) \cdot \pi(\eta_0),
\end{multline*}
where $f(y_{ij}\mid n_{ij}, \theta_{ij}) = [e^{\theta_{ij}} / (1 + e^{\theta_{ij}})]^{y_{ij}} \cdot [1 / (1 + e^{\theta_{ij}})]^{n_{ij} - y_{ij}}$, and $\pi(\cdot)$ represents the corresponding prior densities as in Equation \eqref{eq:bhm}.
Posterior samples of the unknown parameters, 
\begin{align*}
\{\bm{\theta}^{(r)},\bm{Z}^{(r)},\bm{\xi}^{(r)},\bm{\eta}^{(r)},{\xi_0}^{(r)},{\eta_0}^{(r)}; r = 1, \ldots, R\}, 
\end{align*}
are obtained through Markov chain Monte Carlo (MCMC) simulation, where $R$ denotes the maximum number of MCMC iterations. 
The MCMC simulation follows standard Gibbs and Metropolis-Hastings steps, the detail of which is omitted.

\subsection{Proposed Trial Design}

Based on the probability model in Section \ref{sec:prob_model}, we propose the MUCE design for multiple expansion cohort trials. 
The MUCE design without interim looks can be derived based on the following logic.
We enroll $n_{ij}$ patients to arm $(i, j)$, and declare the arm promising if
\begin{align*}
\Pr(\lambda_{ij} = 1 \mid \mathcal{D}) > \phi_2
\end{align*}
or not promising if $\Pr(\lambda_{ij} = 1 \mid \mathcal{D}) < \phi_1$. 
Here, $\mathcal{D} = \{ (n_{ij}, y_{ij}); i = 1, \ldots, I, j = 1, \ldots, J \}$ denotes the observed data at the end of the trial, where $y_{ij}$ is the number of responders in arm $(i, j)$. The posterior probability of $H_{1,ij}$ being true (i.e., $\lambda_{ij} = 1$) can be approximated from the posterior MCMC samples,
\begin{align*}
\Pr(\lambda_{ij}=1 \mid \mathcal{D}) \approx \frac{1}{R} \sum_{r=1}^R  I(Z_{ij}^{(r)}\geq 0).
\end{align*} 
Recall that $\{ Z_{ij}^{(r)}; r = 1, \ldots, R \}$ denotes $R$ posterior samples of $Z_{ij}$.
From a Bayesian perspective, cutoff $\phi_2$ is specified so that $(1-\phi_2)$ gives a desired posterior probability of null (PPN) when arm $(i,j)$ is considered promising, i.e., a false positive decision is made. For example, $\phi_2=0.9$ gives a PPN of 0.1 as the upper bound for making a false positive decision using the Bayesian inference. Similarly, the value of $\phi_1$ provides the upper bound of the posterior probability of alternative (PPA). For example, $\phi_1 = 0.3$ gives a small PPA and indicates a small probability of making a false negative decision given the data and the MUCE model.
After $\phi_2$ and $\phi_1$ are specified, the sample sizes $\{ n_{ij}; i = 1, \ldots, I, j = 1, \ldots, J \}$ are decided based on simulation so that desirable frequentist type I/II error rates are achieved.

Once $\phi_2$, $\phi_1$ and $\{ n_{ij} \}$ are decided, one can add futility interim looks to the MUCE design. Suppose $L (\geq 1)$ interim looks are planned, and interim analysis $l$ is conducted after $n_{ij}^l$ patients have been enrolled in arm $(i, j)$, where $n_{ij}^l < n_{ij}$. 
Let $\mathcal{D}^l = \{ (n_{ij}^l, y_{ij}^l); i = 1, \ldots, I, j = 1, \ldots, J \}$ denote the observed data at interim analysis $l$, where $y_{ij}^l$ is the number of responders among the $n_{ij}^l$ patients. At each interim analysis, arm $(i, j)$ is stopped early for futility if $\Pr(\lambda_{ij} = 1 \mid \mathcal{D}^l) < \phi_1$. See Box 1. 
Note that the maximum sample sizes $\{ n_{ij} \}$ may be recalibrated if interim looks are planned, again based on simulation.

\begin{center}
\fbox{
\parbox{.96\textwidth}
{
\textbf{Box 1: The MUCE design with $L$ futility interim looks. } 
\vskip -.1in
\begin{enumerate}[leftmargin=*, label*=\arabic*., start=0]
\item Let $l=1$.
\item After $n_{ij}^l$ patients 
have been enrolled in arm $(i, j)$, calculate $\Pr(\lambda_{ij}=1 \mid \mathcal{D}^l)$.
If $\Pr(\lambda_{ij}=1 \mid \mathcal{D}^l)  <\phi_1$, stop patient accrual in this arm for futility.
\item If patient accrual in all arms has been stopped, stop the trial. Otherwise, let $l=l+1$.
\begin{enumerate}[leftmargin=*, label=(\alph*)]
\item If $l \leq L$, go back to step 1;
\item Otherwise, enroll patients until the maximum sample size $n_{ij}$ is reached for arm $(i, j)$. Evaluate each arm based on the final observed data. If $\Pr(\lambda_{ij}=1 \mid \mathcal{D})>\phi_2$, declare arm $(i,j)$ promising at the end of the trial.
\end{enumerate} 
\end{enumerate}
}}
\end{center}

\section{Results}\label{sec:result}

\subsection{Two Trial Examples}\label{sec:example_simulate}

In this section, we  illustrate the application of the MUCE design through two hypothetical trials, denoted as trial examples I and II.
These examples are based on a simplified version of the motivating example described in Section \ref{sec:intro}. In both examples, one dose is expanded in four indications  (i.e., $J = 1$ and $I = 4$), with the reference response rate $\pi_{i0} = 20\%$  for all indications.
We set $\phi_1 = 0.3$ as the threshold for futility stopping at each interim analysis and $\phi_2 = 0.9$ for declaring treatment efficacy at the final analysis. Recall that $\phi_1$ represents the upper bound of the PPA when a negative decision is made, and $(1 - \phi_2)$ represents the upper bound of the PPN when a positive decision is made.
For simplicity, we set the maximum sample size to be 29 per arm, which is chosen to match the sample size of the Simon's two-stage design with a type I error rate of 0.1, a type II error rate of 0.3, a reference response rate of $\pi_{i0} = 20\%$, and a target response rate of $\pi_{i1} = 35\%$. In practice, the maximum sample size may be chosen based on simulation to attain desirable frequentist properties. 
Two interim looks for futility stopping are conducted after the responses of 10 and 20 patients have been evaluated in every arm, respectively. 
Through these two trial examples, we will show the effect of borrowing across arms and the benefit of  futility stopping.

We apply MUCE under the following three hyperparameter settings:
\begin{enumerate}[label*=(\roman*), noitemsep]
\item Setting 1: $\gamma=2.5$, $\mu_{\xi_0}=\mu_{\eta_0}=0$, and $\sigma_{0}^2=\sigma_{\xi}^2=\sigma_{\eta}^2=\sigma_{\xi_0}^2=\sigma_{\eta_0}^2=1$;
\item Setting 2: Same as Setting 1 except $\sigma_{\xi_0}^2=\sigma_{\eta_0}^2=3^2$;
\item Setting 3: Same as Setting 1 except  $\mu_{\xi_0}=\mu_{\eta_0}=-3$.
\end{enumerate}
Here, Setting 1 is the default hyperparameter setting. Setting 2 imposes more information borrowing compared to Setting 1, as it increases the correlation of $Z_{ij}$ across different arms (see Equation \ref{eq:corr}). Setting 3 places a lower prior probability for $H_{1, ij}$, which makes it easier to stop an arm early due to futility and more difficult to declare treatment efficacy at the end of the trial. 

\begin{figure}[h!]
\centering \includegraphics[width=0.5\textwidth]{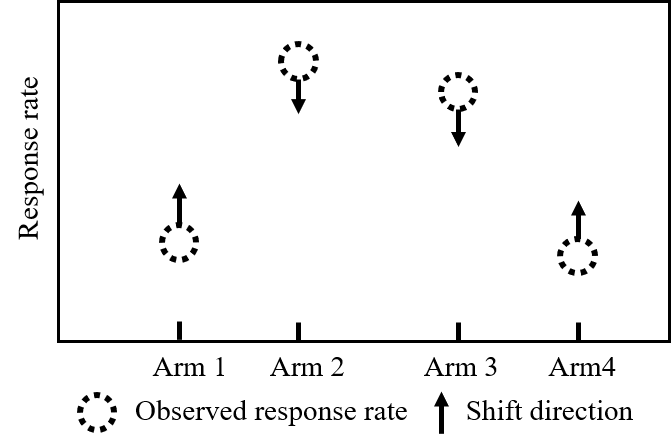}
\caption{An illustration of the effect of information borrowing under MUCE. 
The dotted circles represent the observed/raw response rates in the four arms. 
The solid arrows show the shrinkage direction of the estimated response rates based on MUCE. 
In this example, the estimated response rates in the four arms are shrunk toward the overall mean, with smaller observed response rates in arms 1 and 4 up-shifted and those in arms 2 and 3 down-shifted.} 
\label{fig:shrinkage}
\end{figure}

Table \ref{tab:example1} shows the simulated data 
for trial example I and inference based on MUCE 
under the three hyperparameter settings.
At the first interim look, respectively 1, 5, 6, and 3 responders are reported in the four arms among the first 10 enrolled patients.
Under Setting 1, the posterior probability of $H_{1,ij}$ is greater than $\phi_1 = 0.3$ for all arms, and therefore patient accrual continues in all arms.
The estimated response rates under MUCE show the effect of ``borrowing'', as the smaller observed response rates in arms 1 and 4 are up-shifted and those in arms 2 and 3 are down-shifted. See Figure \ref{fig:shrinkage} for an illustration. 
Setting 2 leads to stronger borrowing, which can be seen from the greater degree of shrinkage of the estimated response rates compared to that under Setting 1.
Again, no arm is stopped early for futility. 
Setting 3 leads to lower estimated response rates and posterior probabilities of $H_{1, ij}$'s, because it assumes a lower prior probability of $H_{1, ij}$ by imposing negative $\mu_{\xi_0}$ and $\mu_{\eta_0}$ values in the latent probit regression. As a result, the posterior probabilities of $H_{1, ij}$'s are also lower, and arm 1 is stopped early due to futility. In other words, negative values of $\mu_{\xi_0}$ and $\mu_{\eta_0}$ lead to more conservative decisions.

\begin{table}
\caption{\label{tab:example1} Trial example I under the MUCE design. ``Est. $p$'' denotes the estimated response rate, and ``Prob. $H_1$'' denotes the posterior probability of the alternative hypothesis, i.e., $\Pr(\lambda_{ij}=1 \mid \mathcal{D})$. The bold values indicate the arms that are declared promising at the final analysis. The values in square brackets indicate the arms that are stopped early due to futility. The values in parentheses indicate that the interim data are carried forward for the subsequent analyses after the arms are stopped early. } 
\centering
\resizebox{\linewidth}{!}{
\begin{tabular}{cc|cccc|cccc|cccc}
\hline\hline
&& \multicolumn{4}{c|}{Interim 1} & \multicolumn{4}{c|}{Interim 2} & \multicolumn{4}{c}{Final Analysis}\\
\hline
\multicolumn{2}{c|}{arm}& 1 & 2& 3 & 4& 1 & 2& 3 & 4& 1 & 2& 3 & 4\\
\multicolumn{2}{c|}{$n$}& 10 & 10& 10 & 10& 20 & 20& 20 & 20& 29 & 29& 29 & 29\\
\multicolumn{2}{c|}{$y$}& 1 & 5& 6 & 3& 4 & 10& 9 & 8& 6 & 13& 11 & 10\\ \hline
\multicolumn{2}{c|}{$y/n$}& 0.1 & 0.5& 0.6 & 0.3& 0.2 & 0.5& 0.45 & 0.4& 0.21 &0.45 &0.38 &0.34\\\hline
    \multirow{3}[2]{*}{Est. $p$} & Set. 1 & 0.139 & 0.473 & 0.572 & 0.304 & 0.240  & 0.483 & 0.435 & 0.388 & 0.237 & 0.437 & 0.371 & 0.340 \\
    & Set. 2 & 0.199 & 0.478 & 0.574 & 0.326 & 0.258 & 0.482 & 0.437 & 0.387 & 0.254 & 0.438 & 0.368 & 0.340 \\
    & Set. 3 & [0.076] & 0.361 & 0.462 & 0.198 & (0.082) & 0.479 & 0.426 & 0.370  & (0.084) & 0.431 & 0.355 & 0.314 \\
    \hline
   \multirow{3}[2]{*}{Prob. $H_1$} & Set. 1 & 0.482 & 0.987 & 0.997 & 0.862 & 0.814  & 0.999  & 0.998  & 0.993  & 0.828  & \textbf{1.000 } & \textbf{0.995 } & \textbf{0.987 } \\
   & Set. 2 & 0.747 & 0.994 & 0.999 & 0.944 & 0.930  & 1.000  & 0.999  & 0.997  & \textbf{0.945 } & \textbf{1.000 } & \textbf{0.999 } & \textbf{0.997 } \\
   & Set. 3 & [0.076] & 0.687 & 0.762 & 0.370  & (0.144) & 0.987  & 0.969  & 0.923  & (0.130) & \textbf{0.977 } & \textbf{0.932 } & 0.864  \\
   \hline\hline
\end{tabular}}
\end{table}

The second interim analysis occurs after 20 patients have been assessed for response in every arm. Again, under Setting 1, the futility stopping boundary is not crossed, and the trial continues with all four arms. At the end of the trial, 6, 13, 11 and 20 responders are observed in arms 1, 2, 3 and 4, respectively. The posterior probabilities of $H_{1, ij}$ for arms 2, 3 and 4 are over the efficacy threshold $\phi_2 = 0.9$, and the dose is considered promising in these arms.
Under Setting 2, the posterior probability of $H_{1, ij}$ is higher in all arms due to stronger borrowing compared to that under Setting 1, and the dose is considered promising in all arms. 
Under Setting 3, due to the lower prior probability of $H_{1, ij}$, the posterior probability of $H_{1, ij}$ is lower in all arms, and the dose is considered promising only in arms 2 and 3. 
Hyperparameter Setting 3 may be chosen if the investigators place strong emphasis on type I error control. 
Note that under Setting 3, although patient accrual in arm 1 is stopped after the first interim analysis,  the interim data for arm 1 (1 responder out of 10 patients) are still included in the second interim and the final analyses, a benefit of Bayesian modeling.

Table \ref{tab:example2} presents the second trial example. 
At the first interim analysis, 0, 3, 6 and 4 responders are observed in arms 1, 2, 3 and 4, respectively.
The posterior probability of $H_{1, ij}$ is only 0.046 for arm 1 under Setting 1.  As a result, arm 1 is stopped for futility. We can see similar performance of the MUCE design in the second trial example: Setting 2 has stronger borrowing strength than Setting 1, and Setting 3 has strong type I error control. 
Notice that due to early stopping of arm 1 under all three settings, we did not simulate any additional data for arm 1 in the second and final analysis. The arm is terminated after interim look 1 to avoid treating more patients in this potentially non-promising arm. Under Setting 3,  arm 2 is also terminated after interim look 1. At the end of the trial, the posterior probabilities of $H_{1, ij}$ for arms 2, 3 and 4 are over the efficacy threshold $\phi_2 = 0.9$ under Settings 1 and 2. For Setting 3, only arms 3 and 4 are declared promising due to the strong type I error control.

\begin{table}
\caption{\label{tab:example2} Trial example II under the MUCE design. ``Est. $p$'' denotes the estimated response rate, and ``Prob. $H_1$'' denotes the posterior probability of the alternative hypothesis. The bold values indicate the arms that are declared promising at the final analysis. The values in square brackets indicate the arms that are stopped early due to futility. The values in parentheses indicate that the interim data are carried forward for the subsequent analyses after the arms are stopped early.}
\centering
\resizebox{\linewidth}{!}{
\begin{tabular}{cc|cccc|cccc|cccc}
\hline\hline
&& \multicolumn{4}{c|}{Interim 1} & \multicolumn{4}{c|}{Interim 2} & \multicolumn{4}{c}{Final Analysis}\\
\hline
\multicolumn{2}{c|}{arm}& 1 & 2& 3 & 4& 1 & 2& 3 & 4& 1 & 2& 3 & 4\\
\multicolumn{2}{c|}{$n$}& 10 & 10& 10 & 10& 10 & 20& 20 & 20& 10 & 29& 29 & 29\\
\multicolumn{2}{c|}{$y$}& 0 & 3& 6 & 4& 0 & 6& 10 & 8& 0 & 9& 14 & 11\\ \hline
\multicolumn{2}{c|}{$y/n$}& 0.0 & 0.3& 0.6 & 0.4& 0.0 & 0.3& 0.5 & 0.4& 0.00& 0.31& 0.48& 0.38\\\hline
\multirow{3}[2]{*}{Est. $p$} & Set. 1 & [0.002] & 0.286  & 0.574  & 0.371  & (0.000) & 0.295  & 0.484  & 0.385  & (0.004) & 0.303  & 0.471  & 0.369  \\
& Set. 2 & [0.003] & 0.298  & 0.572  & 0.382  & (0.001) & 0.299  & 0.484  & 0.386  & (0.006) & 0.307  & 0.473  & 0.369  \\
& Set. 3 & [0.000] & [0.171] & 0.397  & 0.220  & (0.000) & (0.235) & 0.464  & 0.343  & (0.000) & (0.272) & 0.470  & 0.353  \\
\hline
\multirow{3}[2]{*}{Prob. $H_1$} & Set. 1 & [0.069] & 0.800  & 0.996  & 0.918  & (0.059) & 0.887  & 0.999  & 0.980  & (0.084) & \textbf{0.936 } & \textbf{0.999 } & \textbf{0.988 } \\
& Set. 2 & [0.184] & 0.840  & 0.996  & 0.940  & (0.153) & 0.910  & 0.998  & 0.983  & (0.229) & \textbf{0.956 } & \textbf{1.000 } & \textbf{0.992 } \\
& Set. 3 & [0.004] & [0.233] & 0.620  & 0.366  & (0.010) & (0.550) & 0.941  & 0.823  & (0.003) & (0.749) & \textbf{0.996 } & \textbf{0.924 } \\
\hline\hline
\end{tabular}}
\end{table}

We can also observe the effect of borrowing strength across arms by comparing the two trial examples. For example, arm 3 in trial example I and arm 4 in trial example II have exactly the same observed data (29 patients in total with 11 responders), while inference about $H_{1, ij}$ for these two arms is slightly different in the two trial examples. This is because such inference is affected by the observed data in the other arms,  which are different between the two trials.

\subsection{Simulation 1: One Dose and Multiple Indications}
\label{sec:sim1}

We conduct extensive simulations to examine the operating characteristics of the proposed MUCE design.
In the first simulation study, we aim to benchmark the performance of MUCE against the Simon's two-stage design in terms of frequentist power, type I error rate, and average sample size.
We also include the BBHM, EXNEX and CBHM designs in the comparison. 
For a fair comparison, we only consider one dose and four indications (i.e., $J = 1$ and $I = 4$), since BBHM, EXNEX and CBHM are developed for basket trials rather than expansion cohort trials with two factors: doses and indications.

We consider five different scenarios, shown in Table \ref{tab:scenario1}.
We assume the reference response rate is $\pi_{i0} = 0.2$ for all indications. 
We also set the target response rate $\pi_{i1} = 0.35$, which is required for implementing the Simon's two-stage, BBHM, EXNEX and CBHM designs.
Under each scenario, patient responses are generated according to the true response rates. The first scenario is a global null scenario, in which all arms are non-promising having a response rate of 0.2. 
The second scenario is a global alternative scenario with all arms promising having a response rate of 0.35. 
Scenarios 3--5 are mixed scenarios, with different numbers of promising and non-promising arms.

\begin{table}
\caption{\label{tab:scenario1}True response rates of the four arms (indications) under the five scenarios in Simulation 1. The bold values mark the promising arms. }
\centering
\begin{tabular}{c|cccc}
\hline\hline
Scenario & arm 1 & arm 2&arm 3&arm 4\\\hline
1 &0.2& 0.2& 0.2& 0.2\\
2&\bf 0.35& \bf0.35& \bf0.35& \bf0.35\\
3&0.2& 0.2& \bf0.35& \bf0.45\\
4&0.2&\bf0.35&\bf0.35&\bf0.45\\
5&0.2&0.2&0.2&\bf0.35\\
\hline\hline
\end{tabular}
\end{table}

The Simon's two-stage design with a prespecified type I error rate of 0.1 and a type II error rate of 0.3 is given by the following:  
for each arm, treat 13 patients in the first stage. If $\leq 2$ patients respond, stop the arm early; otherwise, treat additional 16 patients in the second stage (29 patients in total), and declare the arm promising if $> 8$ patients respond in total.
To match the maximum sample size of the Simon's two-stage design, the maximum sample sizes for MUCE, BBHM, EXNEX and CBHM are also set at 29 for every arm.
Two interim looks for futility stopping are conducted after 10 and 20 patient outcomes are observed in every arm for these four Bayesian designs. 
The MUCE design is implemented under hyperparameter Setting 1 (see Section \ref{sec:example_simulate}). The futility stopping boundary and the efficacy thresholds are chosen as $\phi_1^{\text{MUCE}} = 0.25$ and $\phi_2^{\text{MUCE}} = 0.924$, respectively. 
The BBHM, EXNEX and CBHM designs are implemented under the default hyperparameter settings recommended in the corresponding publications. 
The futility and efficacy thresholds for BBHM, EXNEX and CBHM are set at $\phi_1^{\text{BBHM}} = \phi_1^{\text{EXNEX}} = \phi_1^{\text{CBHM}} = 0.08$,  $\phi_2^{\text{BBHM}} = 0.879$,  $\phi_2^{\text{EXNEX}} = 0.950$ and $\phi_2^{\text{CBHM}} = 0.957$, respectively.
These thresholds are chosen such that (i) all designs yield approximately the same average sample size ($\approx 21$) under the global null scenario (Scenario 1), and (ii) all Bayesian designs have the same family-wise type I error rate (FWER) ($= 0.15$) under the global null scenario.
Here, a  family-wise type I error refers to a decision in which at least one non-promising arm is falsely declared to be promising (i.e., at least one true null hypothesis is rejected).
The purpose of calibrating the threshold values is to benchmark the comparison among different designs.

\begin{figure}[h!]
\begin{center}
\includegraphics[width=\textwidth]{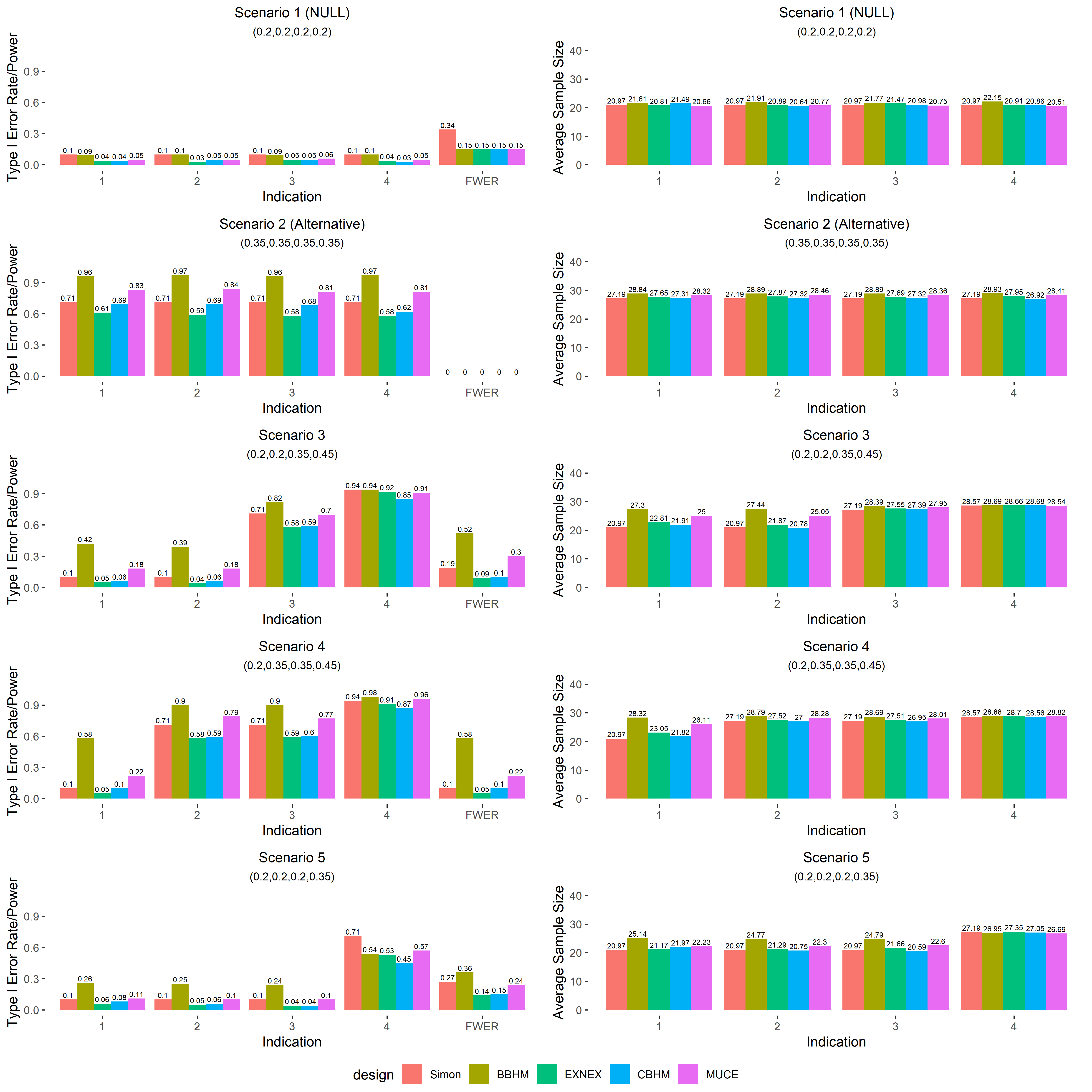}
\caption{Comparison of power, type I error rate (left panel), and average sample size (right panel) of the Simon's two-stage, BBHM, EXNEX, CBHM, and MUCE designs under the five scenarios in Simulation 1 (all with one dose level and four indications).} 
\label{fig:simul1}
\end{center}
\end{figure}

We simulate 1,000 trials under each scenario (Table \ref{tab:scenario1}) for each design.  
We record (i) the percentage of trials in which an arm is declared promising. This is the type I error rate if the arm is actually non-promising, or the power if the arm is truly promising. In addition to (i), we also record (ii) the percentage of trials in which any non-promising arm is falsely declared promising, i.e., the FWER, and (iii) the average sample size.
The simulation results are shown in Figure \ref{fig:simul1}. 
In Scenario 1, although the arm-wise type I error rate of the Simon's two-stage design is controlled at 0.1,  it has a FWER of 0.34. All the four Bayesian designs have arm-wise and family-wise type I error rates lower than those of the Simon's two-stage design.
In Scenario 2, BBHM has the highest power in all arms, followed by MUCE, Simon's two-stage, CBHM and EXNEX.
The high power of BBHM in Scenario 2 is attributed to its strong borrowing of strength across arms, as shown in \cite{berry2013bayesian}.
In the mixed scenarios (Scenarios 3--5), the Simon's two-stage design is able to control the type I error rates for the non-promising arms at 0.1, because inference for each arm is conducted separately. The BBHM design has elevated type I error rates in the non-promising arms due to its strong borrowing behavior. We can also observe some type I error inflation for the MUCE design, but such inflation is less extreme compared to the BBHM design and is considered reasonable.
Given that MUCE is not designed for basket trials, its performance exhibited in this simulation seems satisfactory. 
In summary, MUCE is able to
\begin{enumerate}[label*=(\roman*), noitemsep]
\item Control arm-wise and family-wise type I error rates under the global null scenario,
\item Exhibit desirable power under the global alternative scenario, and
\item Strike a good balance between type I error rate and power under the mixed scenarios. That is, MUCE shows sufficient power in selecting the promising arms without greatly inflating the type I error rate in selecting the non-promising arms.
\end{enumerate}
The average sample sizes of the five designs are also reported in Figure \ref{fig:simul1}, which are generally similar, although the Simon's two-stage design has slightly lower average sample sizes in some cases.

\subsection{Simulation 2: Multiple Doses and Multiple Indications}
\label{sec:sim2}

In the second simulation study, we consider the motivating 
phase 1b multiple expansion cohort trial example described in Section \ref{sec:intro}.
Suppose three doses are graduated from phase 1a dose-escalation to phase 1b expansion cohort, and four indications are of interest (i.e., $J = 3$ and $I = 4$).
As a result, 12 different dose-indication arms are available for expansion. 
The trial budget only allows a total sample size of 120 patients with 10 patients per arm.
We conduct simulation to examine the frequentist operating characteristics as part of the initial new drug (IND) application to the regulatory agency.

We consider six scenarios (Table \ref{tab:scenario2}) that specify the true response rates of the 12 arms.
We assume the reference response rate is $\pi_{i0} = 0.2$ for all the four indications.
The first scenario is a global null scenario with all arms non-promising having a response rate of 0.2. The second scenario is a global alternative scenario with all arms promising having a response rate of 0.5. This value is considered clinically beneficial to patients, and an arm exhibiting such a response rate deserves further clinical development. 
Scenario 3 is also a global alternative scenario, in which all the arms have response rates higher than 0.2 but ranged from 0.3 to 0.5. 
Scenarios 4--6 are mixed scenarios with promising and non-promising arms. The promising arms in Scenarios 4 and 6 have a response rate of 0.5 regardless of the dose, and the promising arms in Scenario 5 show an increasing dose-response trend.

\begin{table}
\caption{\label{tab:scenario2} True response rates of the twelve arms under the six scenarios in Simulation 2. The bold values mark the promising arms.}
\centering
\begin{tabular}{c|c|cccc}
\hline\hline
Scenario & dose level & indication 1 & indication 2&indication 3&indication 4\\\hline
\multirow{3}{*}{1} &1 & 0.2& 0.2& 0.2& 0.2\\
&2 & 0.2& 0.2& 0.2& 0.2\\
&3 & 0.2& 0.2& 0.2& 0.2\\\hline
\multirow{3}{*}{2} &1 & \bf 0.5& \bf 0.5& \bf 0.5& \bf 0.5\\
&2 & \bf 0.5& \bf 0.5& \bf 0.5& \bf 0.5\\
&3 & \bf 0.5& \bf 0.5& \bf 0.5& \bf 0.5\\\hline
\multirow{3}{*}{3} &1 & \bf 0.3& \bf 0.3& \bf 0.3& \bf 0.3\\
&2 & \bf 0.4& \bf 0.4& \bf 0.4& \bf 0.4\\
&3 & \bf 0.5& \bf 0.5& \bf 0.5& \bf 0.5\\\hline
\multirow{3}{*}{4} &1 & \bf 0.5& \bf 0.5& 0.2& 0.2\\
&2 & \bf 0.5& \bf 0.5& 0.2& 0.2\\
&3 & \bf 0.5& \bf 0.5& 0.2& 0.2\\\hline
\multirow{3}{*}{5} &1 & \bf 0.3& \bf 0.3& 0.2& 0.2\\
&2 & \bf 0.4& \bf 0.4& 0.2& 0.2\\
&3 & \bf 0.5& \bf 0.5& 0.2& 0.2\\ \hline
 \multirow{3}{*}{6} &1 & 0.2 & 0.2 & 0.2& 0.2\\
&2 & 0.2 & 0.2 & 0.2& 0.2\\
&3 & \bf 0.5& \bf 0.5& 0.2& 0.2\\
\hline\hline
\end{tabular}
\end{table}

We assess the performance of the MUCE design under the default hyperparameter Setting 1 and compare it with that of the BBHM, EXNEX and CBHM designs.
The Simon's two-stage design is not considered here since it will lead to unacceptable FWER with 12 arms.
We simulate 1,000 trials under each scenario (Table \ref{tab:scenario2}) for each design.
With 10 patients per arm, no interim look is implemented for all designs. Therefore, we do not need to specify the target response rate for BBHM, EXNEX and CBHM, which is only used for interim futility stopping.
For a fair comparison, the efficacy thresholds $\phi_2$ for these four methods are calibrated to generate an identical FWER of 0.1 under Scenario 1 (global null). 
We obtain $\phi_2^{\text{MUCE}} = 0.988$, $\phi_2^{\text{BBHM}} = 0.948$,  $\phi_2^{\text{EXNEX}} = 0.976$ and $\phi_2^{\text{CBHM}} = 0.989$.

\begin{figure}[h!]
\begin{center}
\includegraphics[width=\textwidth]{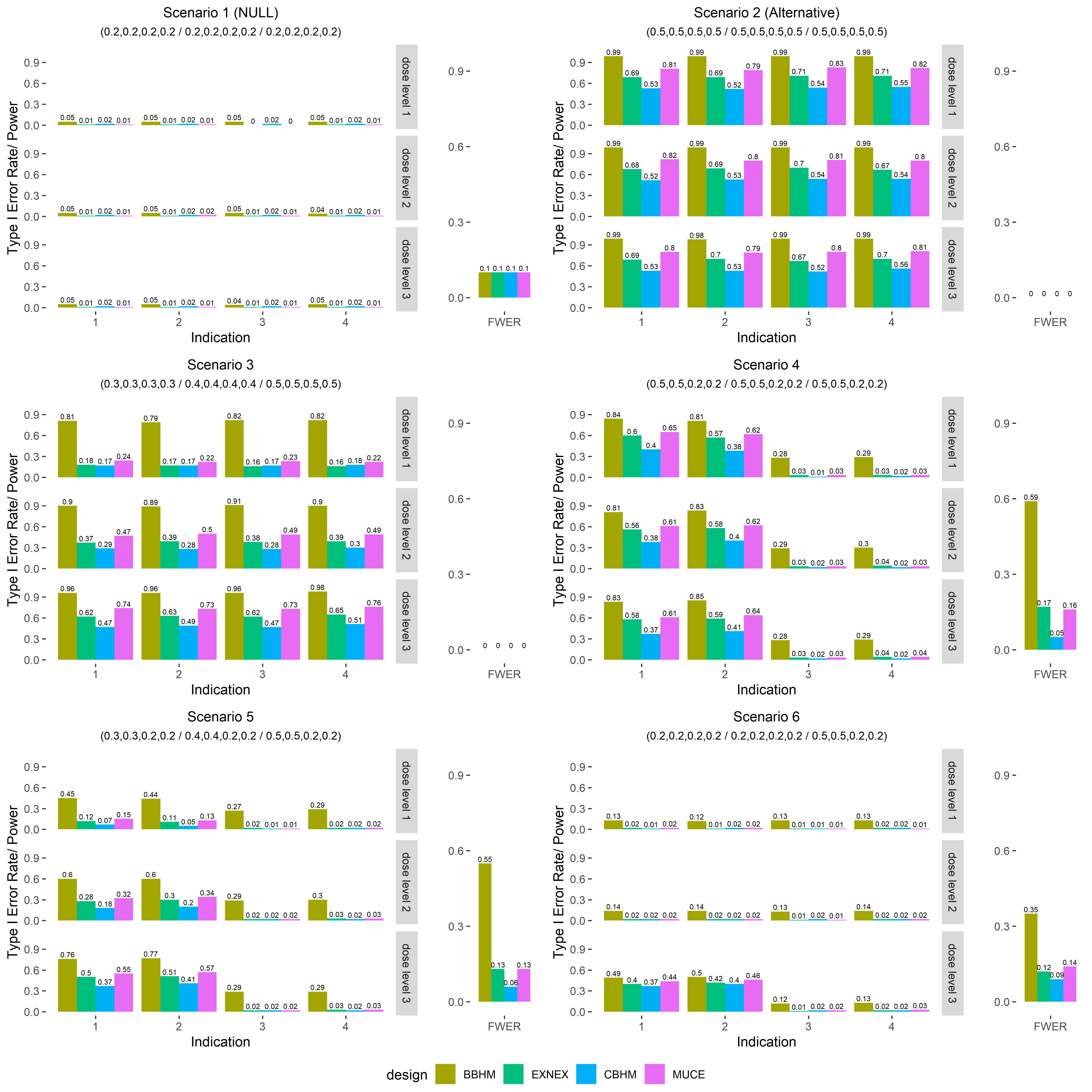}
\caption{Comparison of power and arm-wise and family-wise type I error rates of the BBHM, EXNEX, CBHM, and MUCE designs under the six scenarios in Simulation 2 (all with three dose levels and four indications).}
 \label{fig:simul2} 
\end{center}
\end{figure}

Figure \ref{fig:simul2} shows the power, arm-wise, and family-wise type I error rates of the different designs under the six scenarios. 
In Scenario 1, all designs have the same FWER of 0.1 because of the threshold calibration.
In Scenarios 2 and 3, BBHM has the highest power to detect the promising arms, followed by MUCE $\geq$ EXNEX $>$ CBHM. This is expected, as BBHM has the highest degree of borrowing, which allows it to perform better in the global alternative scenarios.
In the mixed scenarios (Scenarios 4--6),  although BBHM still has the highest power for detecting the promising arms, it shows inflated arm-wise and family-wise type I error rates. Furthermore, MUCE generally has better power and type I error control compared to EXNEX. Lastly, CBHM has the best type I error control but lacks sufficient power to detect the promising arms.
The reason why MUCE has both decent power and type I error control is that it exploits the two-way expansion data structure and employs a latent probit model that allows different degrees of borrowing across doses and indications.
In contrast, BBHM, EXNEX and CBHM only consider one-dimensional information borrowing.

\subsection{Sensitivity Analysis and Multiplicity Control} 
\label{sec:sens}

In Section \ref{sec:example_simulate}, we have demonstrated the behavior of the MUCE design under three hyperparameter settings through two trial examples. 
In this section, we conduct sensitivity analysis to assess the frequentist operating characteristics of MUCE under more hyperparameter settings and investigate the effect of different hyperparameters. 
In addition to hyperparameter Settings 1--3 in Section \ref{sec:example_simulate}, we consider two more hyperparameter settings:
\begin{enumerate}[label*=(\roman*), noitemsep]\addtocounter{enumi}{3}
\item Setting 4: Same as Setting 1 except $\sigma_{\xi_0}^2=\sigma_{\eta_0}^2=0.1^2$;
\item Setting 5: Same as Setting 1 except $\mu_{\xi_0} = -3$.
\end{enumerate}
Setting 4 imposes weaker borrowing across arms than Setting 1, as it decreases the correlation of $Z_{ij}$ across arms (see Equation \ref{eq:corr}).
Setting 5 provides weaker multiplicity control compared to Setting 3, although it still has stronger multiplicity control than Setting 1.

We consider simulation Scenarios 1--3 in Table \ref{tab:scenario1} with one dose level and four indications.
For each scenario, we simulate 1,000 trials with the MUCE design under each hyperparameter setting. 
Again, we set the maximum sample size for each arm at 29.  For simplicity, we do not implement interim looks for futility stopping during the trial. 
At the end of the trial, the threshold for declaring treatment efficacy is $\phi_2 = 0.95$ for every hyperparameter setting. 

\begin{figure}[h!]
\begin{center}
\includegraphics[width=.9\textwidth]{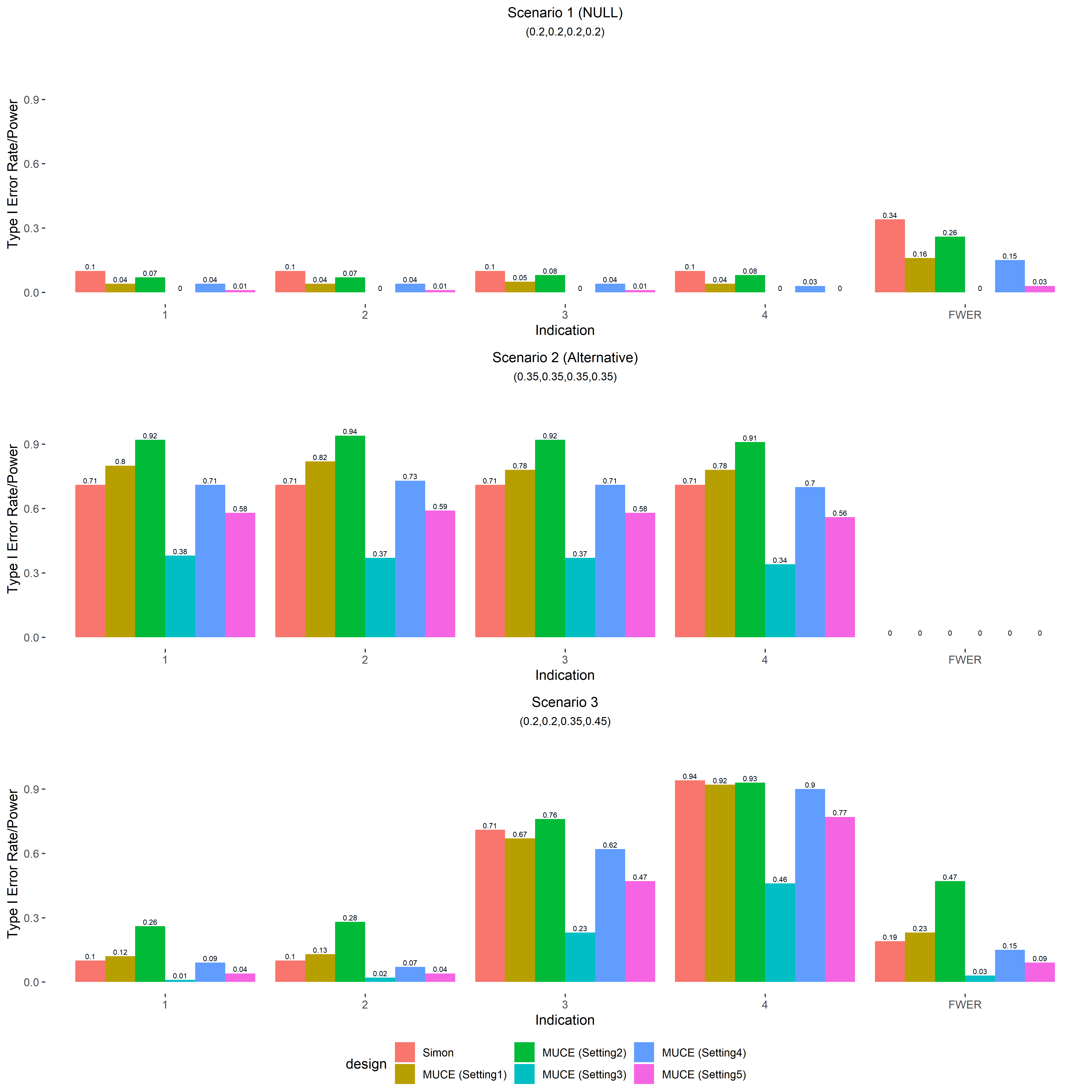}
\caption{Comparison of operating characteristics of the MUCE design under five different hyperparameter settings. The results are benchmarked with the Simon's two-stage design.} 
\label{fig:simul_sensitivity}
\end{center}
\end{figure}

The frequentist type I error rates and powers of MUCE under different hyperparameter settings are reported in Figure \ref{fig:simul_sensitivity}. 
The results using the Simon's two-stage design are also included in Figure \ref{fig:simul_sensitivity} as a benchmark. 
The FWERs of MUCE under Settings 1, 2 and 4 are around 0.15 in Scenario 1,
which are smaller than that of the Simon's two-stage design.
The two settings with stronger multiplicity control, Settings 3 and 5, lead to much lower FWERs in Scenario 1.
In Scenario 2, the power ordering of Settings 1, 2 and 4 is Setting 2 $>$ Setting 1 $>$ Setting 4, which means that the power in the global alternative scenario increases as the strength of borrowing increases. 
However, the ordering of type I error rate in Scenario 3 among Settings 1, 2 and 4 is also Setting 2 $>$ Setting 1 $>$ Setting 4, meaning that strong borrowing strength leads to inflation of the type I error rate in the mixed scenario.
Because of the multiplicity control, the type I error rates are well controlled under Settings 3 and 5, but the powers under Settings 3 and 5 are also lower than those under the other settings in both Scenarios 2 and 3. 

\section{Discussion}\label{sec:discussion}

We have proposed the MUCE design, which is a new Bayesian design for phase 1b multiple expansion cohort trials. 
We take a formal Bayesian hypothesis testing approach to decide which dose-indication combinations are promising for further investigation.
Priors on the null and alternative hypotheses are constructed, which lead to inference directly based on conditional (posterior) probabilities of the hypotheses.
To adaptively borrow information across arms, we build a latent probit model that allows different degrees of borrowing across doses and indications.
Through simulation studies, we have shown that the MUCE design has desirable operating characteristics and compares favorably to existing designs for multiple expansion cohort trials. 
We have also shown that the degree of borrowing and 
multiplicity control can be adjusted through intuitive hyperparameter tuning.

Elicitation of the prior hyperparameters in the MUCE design can be discussed with the clinical team based on the following two considerations. First, how strongly the team prefers to borrow information across doses. This can be realized by increasing (or decreasing) the variances of $\xi_i$ and $\eta_j$'s, which lead to larger (or smaller) correlations of the latent probit scores. 
Second, how strongly the team prefers to control the type I error rate in the presence of multiple tests. This can be realized by assigning a more negative mean value for $\mu_{\xi_0}$ and $\mu_{\eta_0}$, as shown in Section \ref{sec:sens}.

Bayesian designs like MUCE may improve the efficiency of multi-arm trials by borrowing information across arms, which can ideally lead to improved power to detect a treatment effect with a reduced sample size. 
We note that borrowing may result in inflated type I error rates for the non-promising arms if only part of the arms are truly promising.
In addition, multiplicity issues in multiple expansion cohort trials should be of concern, since multiple decisions are made at the end that would result in further development of multiple doses/indications of the drug. A type I error would lead to future failures and waste of resources.

\bibliographystyle{rss}
\bibliography{ref_MUCE}

\end{document}